\documentstyle[aps,epsfig,twocolumn]{revtex}
\begin{document}
\draft
\flushbottom
\twocolumn[
\hsize\textwidth\columnwidth\hsize\csname @twocolumnfalse\endcsname

\title{Stability and dynamics of free magnetic polarons}
\author{M. J. Calder\'on$^{1,2}$, L. Brey$^1$ and P. B. Littlewood$^3$}
\address{$^1$
Instituto de Ciencia de Materiales (CSIC). Cantoblanco,
28049 Madrid. Spain. \\
$^2$ Departamento de F{\'\i}sica Te\'orica de la Materia Condensada,
Universidad Aut\'onoma de Madrid, 28049 Madrid. Spain. \\
$^3$ Cavendish Laboratory, Cambridge University, Madingley Road,
Cambridge CB3 0HE, UK}
\date{\today}
\maketitle
\tightenlines
\widetext
\advance\leftskip by 57pt
\advance\rightskip by 57pt
\begin{abstract}
The stability and dynamics of a free magnetic polaron are studied by
Monte Carlo simulation of a classical two-dimensional Heisenberg model 
coupled to a single electron.
We compare our results to the earlier mean-field analysis of the
stability of the polaron,
finding qualitative similarity but quantitative differences. 
The dynamical simulations
give estimates of the temperature dependence of the polaron diffusion, 
as well
as a crossover to a tunnelling regime.

\end{abstract}
]
\narrowtext
\tightenlines
Recently, the interest on free magnetic polarons (FMP) has been renewed due to
its relation with colossal magnetoresistance \cite{M-L1} 
in manganese pyrochlores \cite{exp-pyro,fontcuberta} and in double-exchange models of magnetism
in the manganite perovskites \cite{khomskii,dagotto,perovs-pol}. 

Free magnetic polarons are magnetically self-trapped carriers in contrast
to the more common bound magnetic polarons which become trapped by an impurity.
These were extensively
studied in diluted magnetic semiconductors\cite{semimag-review} 
and rare earth chalcogenides\cite{Eu-review}.
Golnik et al. \cite{Golnik} found experimental evidence of the existence of not only
bound but free magnetic polarons in $Cd_{1-x}Mn_xTe$ and $Pb_{1-x}Mn_xTe$. 
Their existence leads to an activated behavior of the resistivity above the Curie
temperature that is found in materials with negative magnetoresistance, as well as a temperature-
dependent spin-splitting seen in magneto-optical experiments\cite{Navrocki}.
Theoretical models of FMP have been developed within a mean-field approach \cite{kasuya,nagaev,benoit} 
and generalized to a fluctuation-dominated regime \cite{spalek}. 
In most of these systems, the underlying (super)-exchange interaction between the localized spins
is antiferromagnetic in nature; we shall however be concerned with the ferromagnetic case.

We study the model of a single electron interacting with a spin
background that itself is ordering ferromagnetically \cite{M-L1}. 
We consider the Hamiltonian:

\begin{eqnarray}
H=- & t &\sum_{\langle i,j \rangle\sigma}c_{i\sigma}^{\dagger}c_{j\sigma} \nonumber\\ 
- & J &\sum_{\langle i,j \rangle}\{(1-\alpha)( S_i^x \cdot S_j^x +
S_i^y \cdot S_j^y) +S_i^z \cdot S_j^z\} \nonumber \\
 -& J'& \sum_i \vec \sigma_i \cdot \vec S_i
\label{hamil}
\end{eqnarray}
where $\vec S_i$ refer to the spin of the magnetic ions in the system.
$c_{i\sigma}^{\dagger}$ creates an electron with spin $\sigma$ on the site $i$,
$\vec \sigma_i = c_{i\alpha}^{\dagger}\vec \sigma_{\alpha \beta}c_{i\beta}$
is the conduction spin operator and $\langle i,j \rangle$ denotes sum over the
nearest-neighbors pairs. We have added to the Heisenberg term a 
small Ising anisotropy $\alpha = 0.1$
to improve convergence at low temperatures and by enforcing a non-zero transition
temperature $T_c$ \cite{serena}.  $J'$ is the coupling between a localized spin and
a conduction electron. 
The qualitative behavior is well understood from previous mean-field analyses
\cite{M-L1,kasuya,benoit}. Below a temperature $T_p$ a ferromagnetic polaron
forms by self-trapping in a ferro-magnetically aligned cluster of spins. As the temperature
is lowered toward the Curie temperature $T_c$ the polaron grows in size and becomes
more stable, because the small-q magnetic susceptibility is growing. Near and below $T_c$
the polaron will again become unstable because of the ease of motion in the
background ferro-magnetic spin alignment. Notice that this is quite different from
the case of {\em anti-}ferromagnetic coupling of spins, where the polaron may remain stable
well below the magnetic ordering temperature.

There are several deficiencies of the mean field treatment. The most pronounced is a continuum
treatment of the spin background where fluctuations are neglected. This approximation is such that
the paramagnetic state leads to a vanishing exchange coupling, so that the electron is bound in
a potential of depth $J' \bar S$, with $\bar S$ the average magnetization inside the polaron.
As $J'/t \rightarrow \infty$, the potential well becomes arbitrarily deep.
This is undoubtedly a severe overestimate. 

In the paramagnet there will always be low energy states
localized in the band tail \cite{K-A} with energies $O(t)$ above the ferromagnetic ground state, even in the
strong coupling limit. Such low energy states are produced by random fluctuations of a few neighboring
spins into near-alignment. But now one must distinguish between a self-trapped polaron
and a localized band-tail state, if indeed such a distinction is appropriate.

In this paper we address the topic by a dynamical simulation of the Hamiltonian of Eq.(1).
We show that polarons may be distinguished (when they exist) by a spectroscopic gap to band-like
states, and that they move diffusively. As temperature is raised, the polaron level moves toward
the band edge, and begins to resonate with states in the band tail, leading to a crossover to
hopping conductivity.

We perform a classical Monte Carlo simulation (MC) in two dimensions on a square lattice of 
localized spins $\vec S_i$ which are
treated as classical rotors characterized by the angles $\theta_i$ and 
$\phi_i$. We use periodic boundary conditions on a two dimensional lattice of size up
to $30 \times 30$. 

We place a single electron in the system in the lowest energy eigenstate of the Hamiltonian
consisting of the first and third terms of Eq (1), using the instantaneous spin configuration
for the classical spins $\vec S_i$. The resulting wave function leads
to a local magnetic field proportional to $ \vert\psi(x,y)\vert^2 $, used in the
next step of the MC spin simulation. 

The standard Metropolis algorithm is used. Randomly chosen sites suffer a random
change of spin orientation. Changes are allowed if the increment in energy
$\Delta E$ is such that the quantity $exp(-\Delta E/KT)$ is smaller than a
random number between $0$ and $1$. $4000$ reorientations per spin were made for an initial
equilibration and $3000$ to calculate averages after each diagonalization. 
Each diagonalization defines our time step. 
Changing the number of spin reorientations between each diagonalization led to no significant
change in either the magnetization or binding energy.
All the quantities are given in units of $J$, the Heisenberg parameter. 
The hopping parameter $t$ is fixed to $100$ (estimated with the mean field
relation $T_c \sim zJS^2$ and the values for the parameters 
expected for the pyrochlores \cite{M-L1})  as we are interested in the
behavior versus $J'/t$ and the temperature $T$. $T_c=1.8$ in these
units. 

This new approach allows us to calculate in a self-consistent way the wave
function and the magnetic polarization over a large range of temperature. 
In particular, we can explore the region around $T_c$ where the mean-field 
treatment fails.

\begin{figure}
\epsfig{file=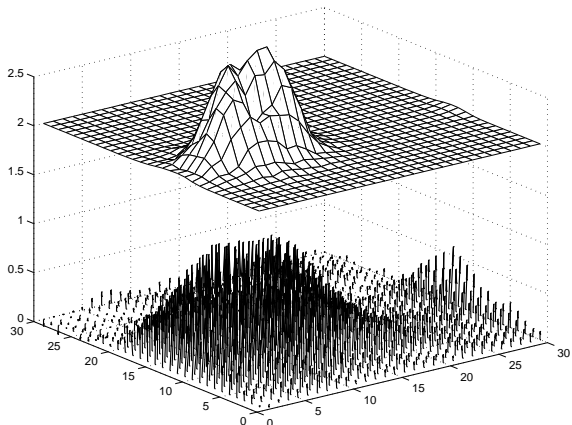,width=3in}
\caption{$\vert\psi(x,y)\vert ^2$ and the averaged magnetization 
$\vec m_i=\frac{\langle \vec S_i \rangle}{\vert \vec S_i \vert}$ for
$J'/t=5$ and $T=1.1T_c$ are plotted.}
\label{fig:wave-function}
\end{figure}

In Fig. \ref{fig:wave-function} we plot $\vert \psi(x,y)\vert ^2$ and the
averaged local magnetization close to $T_c$. Visually, the existence of a magnetic
polaron is clear, and there is substantial alignment of the moments in the vicinity of the
carrier.
Note that far from the
influence of the wave-function the average magnetization is close to 0, 
so a large part of the spin
configurations are explored, while for the spins close to the center of the
wave-function there are few accepted spin flips.

\begin{figure}
\epsfig{file=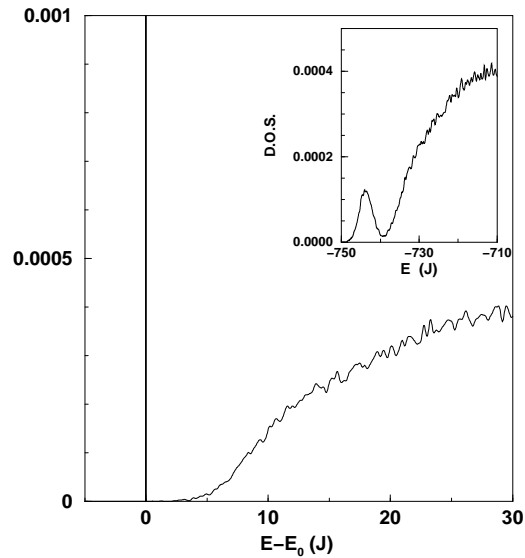,width=2.7in}
\caption{The density of states (inset) and excitation spectrum at low energies
for $J'/t=5$, $T=1.1T_c$ and   $\vert \vec S \vert = 3/2$ is plotted.}
\label{fig:DOS}
\end{figure}

Pictorial evidence is purely qualitative, and does not allow one to extract reliable estimates
for the polaron size or binding energy, especially at higher temperatures and lower $J'/t$, when
the polarons are smaller and fluctuating in time. 
More reliable evidence comes from the time-averaged electronic density of states (DOS), and
of the excitation spectrum shown in Fig. \ref{fig:DOS}. In the
density of states (inset) a sharp low energy feature is pulled from the bottom of the
band (only the lowest 1\% of the spectrum is shown) that contains exactly one state.
This is the bound polaron level. The level width comes from thermal fluctuations
in the energy of the bound state, and the stability of the bound polaron is seen more
clearly in the excitation spectrum (main figure) that demonstrates a clear gap
corresponding to the electronic part of the binding energy $E_p$ of the polaron.

The continuum of excited states can be characterized as the band tail formed by the fluctuating 
paramagnetic background; the lowest energy states are produced by rare fluctuations of nearby
spins into near-alignment. Consequently, the ``gap'' in the excitation spectrum is soft, and
indeed statistically very rare states may occur at energies {\em below} the bound state of the
polaron. We will discuss this below.

\begin{figure}
\epsfig{file=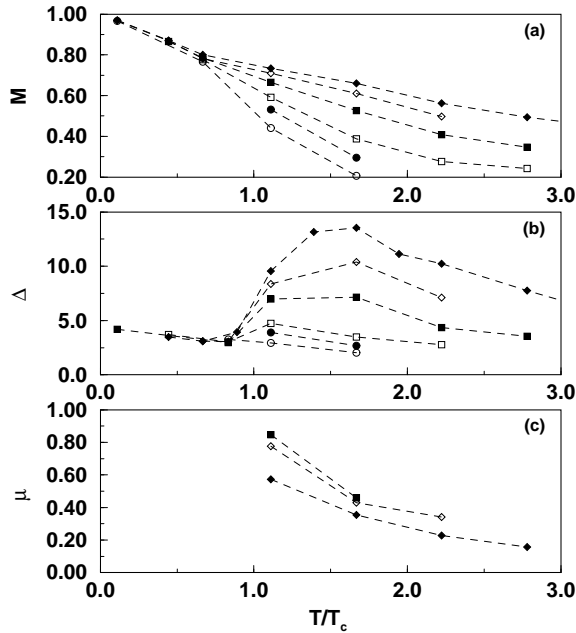,width=3in}
\caption{We plot here the dependence on $J'/t$ and temperature  
of the local magnetization in the lattice produced by the
polaron (a), its binding
energy (b) and the mobility $\mu=\frac{D}{T}$ (c) for $\vert \vec S \vert =3/2$
. The different curves correspond to
different $J'/t$ such that its value is $1$ for open circles, $1.5$ for closed
circles, $2$ for open squares, $3$ for closed squares, $4$ for open diamonds and
$5$ for closed diamonds.}
\label{fig:trends}
\end{figure}

We estimate the electronic binding energy $E_p$ by the configurationally averaged gap 
$\Delta = E_1 - E_0$
to the lowest excited state in our simulations (we have checked that the separation between
excited states scales as $1/N^2$ so that these are true continuum states).
In Fig. \ref{fig:trends} we show the
dependence of $\Delta$ and the absolute value of the local  
magnetization $M$ (weighted with the wave-function) on $T$ and $J'/t$.
$M$ is defined as
\begin{displaymath}
M= \langle \vert  \sum_i \vec S'_i   \vert \rangle
\end{displaymath}
where $ \vec S'_i=\vert \psi(i)\vert ^2 ~ \frac {\vec S_i}{\vert \vec S_i \vert} $.  

We are not taking into
account thermal excitations of the quasiparticle
so our results are valid only when $\Delta$ is
bigger than $T$. This condition is fulfilled for all the values of $\Delta$
shown in Fig. \ref{fig:trends}. 
As expected from previous analyses, the polaron binding energy increases as temperature 
is lowered from
high temperatures, as the thermal spin fluctuations are reduced.
For large $J'/t$ we find a new behavior on
$\Delta$ not found within mean-field theory, namely, that it has a maximum at some temperature above
$T_c$. The existence of a maximum can be 
understood in terms of the correlation length $\xi$. This
quantity increases as we decrease the temperature above $T_c$. For very small
$\xi$ (large $T$) is very difficult to have a FM cluster for the spin-polaron to
sit in and for large $\xi$ the electron would rather spread out. This will lead
to an
intermediate optimum $\xi$ for the existence of the polaron that would happen
close to $T_c$ but not necessarily at $T_c$.

The size of the polaron may be estimated from the separation between the
first eigenvalue $E_o$ and the bottom of the band of the uniform ferromagnet. 
The bottom of the band in this case 
is given by $-\frac{3}{4}J'-4t$ and the separation should go roughly as $\frac
{1}{L_p^2}$ being ${L_p}$ the size of the polaron if we assume saturation in the
local magnetization. The general trend is that 
the size decreases as $T$ (above $T_c$) or $J'$ increases. 
From Fig. \ref{fig:trends} we can also deduce that the 'window' above $T_c$
where the spin-polaron is stable increases with $J'/t$.    
These two results are consistent
with previous mean-field calculations \cite{M-L1}.

Although qualitative comparison is satisfactory there are large quantitative 
differences that point to a great decreasing 
in the stability of the spin-polarons
when fluctuations are taken into account. To be precise we compare the
binding energy at $T=T_c$. From mean-field calculations on ref. \cite{M-L1}
the maximum possible value for $\Delta$ is $\sim J'S$ but it is not reached due
to the kinetic energy that is lost with the formation of a polaron.
In the present
work $\Delta/J \sim 1$ while $J'S/J \sim 100$. So binding energies are reduced 
by two orders of magnitude compared with the mean-field results because the loss
of kinetic energy is not well taken into account in the latter.

\begin{figure}
\epsfig{file=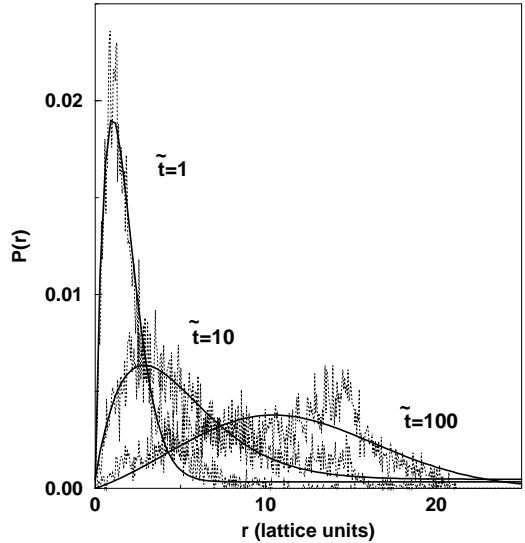,width=2.7in}
\caption{The probability of moving a distance $r$ for different
times, $\tilde t=1$, $10$ and $100$, taken from a single run, is shown. Solid lines
show fits to a 2D Gaussian, plus an offset. This background is due to the rare
appearance of FM clusters away from the polaron location, where the polaron hops a large distance. 
The curves for $\tilde t=10$ and $\tilde t=100$ are the result of iterations of the $\tilde t=1$ curve;
thus the rare long-distance hops eventually dominate the distribution, as can be seen 
in the trace at the longest times.}
\label{fig:prob}
\end{figure}

The study of the stability conditions for a free magnetic polaron is interesting
by itself but the MC simulation also opens us the possibility of learning about
its dynamics in a spin-fluctuating landscape.
In Fig.\ref{fig:prob} the probability of moving a distance $r$ (defined
as the change in the expectation value of the electron position)
for different MC times is shown. For time $\tilde t=1$ one observes dominant short distance
motion with occasional rare hops over long distances.
For longer times, the peak of the distribution moves out approximately with $\sqrt {\tilde t}$ as
expected. This is the expected behavior from a diffusing object.
The long-distance hops occur when unoccupied band tail states (which may be localized
anywhere in the system) temporarily drop below the bound polaron level. In our algorithm -
which automatically populates the lowest energy level - the electron moves to occupy this
new state and restabilises the polaron there. These rare events eventually dominate the
long-time behavior in our simulations. 
Of course, very long range hops are unphysical because the tunnelling
probability will be exponentially small with distance, and the band-tail states survive
in one place for only a short time. Hops to band-tail states will then be limited to some finite
range. As temperature is raised, and $E_p$ is reduced, hops to band tail states become more
frequent; we cross over to a regime of ``passive advection'' of the wavefunction in the
fluctuating spin background \cite{K-A2}.

Our results are fitted to a gaussian in two dimensions plus a
constant (to approximately take account of hops to band-tail states).
The gaussian dominates for the parameters of interest when a spin-polaron is 
well formed. The distribution
scales with $\sqrt {\tilde t}$ as expected for diffusive motion. 
Hence we calculate the diffusion constant as $D=\sum_i P(r,\tilde t=1) r^2$ 
and the mobility
($\mu=\frac{D}{T}$) of the spin polaron for different couplings and temperatures
(see Fig. \ref{fig:trends}). 
The mobility decreases with temperature, and also with $J'/t$. The latter is
reasonable, because larger polarons should diffuse more slowly. The temperature-dependence
is more surprising, and arises because $D$ itself is weakly T-dependent.
Although the polaron size is decreasing with temperature (tending to increase $D$), this
is counterbalanced by a reduced probability of favorable FM spin configurations near its
boundary as $T/T_c$ is increased.

The Heisenberg term has been considered ferromagnetic to compare with the
pyrochlores. The change to antiferromagnetic coupling is straightforward and 
in fact the more common case in manganite perovskites \cite{perovs-pol}, rare earth
chalcogenides \cite {Eu-review}, or magnetic semiconductors \cite{semimag-review}.
We find in the case of an antiferromagnetic background the stability of a free magnetic
polaron is enhanced, and will report these results elsewhere. These results are consistent
with ref. \cite{Moreo} where a pseudogap in the DOS is associated with phase separation,
that is the large scale effect corresponding to spin-polarons.

In conclusion, our dynamical simulations have revealed a picture of the FMP in a ferromagnet above $T_c$
which is considerably more complex than given by
the mean field pictures. Provided the exchange coupling is large enough, FMP's are stable above
$T_c$, but considerably more weakly bound than found by mean field calculations. This by itself raises
some doubts about the interpretation given earlier for the Mn pyrochlores, because we require an 
exchange coupling comparable to the bandwidth for a 
well-formed polaron with nearly saturated magnetization,
whereas in the Mn pyrochlores this coupling is expected to be not large\cite{M-L1}.
We find that the motion of the polaron is diffusive, but as temperature is raised the 
electron fluctuates out of the self-trapped configuration into band-tail states
formed by opportunistic fluctuations of the moments.

{\em Acknowledgments.}
M.J.C. would like to thank  all the TCM group at the Cavendish, 
where the greater part of this work has been done,
for all the fruitful discussions on physics and computing advice. Thanks specially to 
M. C\^ot\'e and L. Wegener. 
P.B.L. aknowledge financial support from EPSRC GR/L55346.
L.B. and M.J.C. also aknowledge financial support from the MEC of Spain under
Contract No. PB96-0085.

\end{document}